\documentclass[a4paper,11pt,showkeys,nofootinbib,longbibliography,superscriptaddress]{revtex4-1}
\usepackage{graphicx}
\usepackage{hyperref,cleveref}  
\usepackage{listings, color}  
\usepackage{amsmath, bm}

\begin{document}

\title{Measurement and comparison of individual external doses of high-school students \\
living in Japan, France, Poland and Belarus\\
-- the ``D-shuttle'' project --}

\author{Adachi,~N.}
\affiliation{Adachi High School, 2-347 Kakunai, Nihonmatsu, Fukushima 964-0904, Japan}
\author{Adamovitch,~V.}
\affiliation{Bragin High School, Bragin, Gomel region, Belarus}
\author{Adjovi,~Y.}
\affiliation{Notre Dame High School, 1 Avenue Charles de Gaulle, 92100 Boulogne-Billancourt, France}
\author{Aida,~K.}
\affiliation{Aizu Gakuho High School,  Ikkimachi Oaza Yahata, Yahata-1-1, Aizuwakamatsu, Fukushima 965-0003, Japan}
\author{Akamatsu,~H.}
\affiliation{Nada High Shool,  8-5-1 Uozakikitamachi, Higashinada-ku, Kobe, Hyogo 658-0082, Japan}
\author{Akiyama,~S.}
\affiliation{Iwaki High School, Taira Aza Takatsuki 7, Iwaki, Fukushima 970-8026, Japan}
\author{Akli,~A.}
\affiliation{Giocante de Casabianca High School, Avenue Jean Zuccarelli, 20200 Bastia, France}
\author{Ando,~A.}
\affiliation{Ena High School, 1023-1 Ohi-cho, Ena, Gifu 509-7201, Japan}
\author{Andrault,~T.}
\affiliation{Bois d'Amour High School, 9 Rue de la Garenne, 86000 Poitiers, France}
\author{Antonietti,~H.}
\affiliation{Notre Dame High School, 1 Avenue Charles de Gaulle, 92100 Boulogne-Billancourt, France}
% ...

\begin{abstract}
\noindent
{ \bf Abstract}

Twelve high schools in Japan (of which six are in Fukushima Prefecture), four in France, eight in Poland and two in Belarus cooperated in the measurement and comparison of individual external doses in 2014.  In total 216 high-school students and teachers participated in the study. Each participant wore an electronic personal dosimeter ``D-shuttle'' for two weeks, and kept a journal of his/her whereabouts and activities. The distributions of annual external doses estimated for each region overlap with each other, demonstrating that the personal external individual doses in locations where residence is currently allowed in Fukushima Prefecture and in Belarus are well within  the range of estimated annual doses due to the {terrestrial} background radiation level of other regions/countries.
\end{abstract}
\keywords{Fukushima Dai-ichi accident, personal dosimetry, international comparison, radiation education}
\maketitle

\section{Introduction}
The Fukushima Dai-ichi nuclear power plant accident, which began  in March 2011, released a significant amount of radioactive substances, contaminating Fukushima and surrounding prefectures~\cite{tanakaaccident2012}. It is therefore essential to clarify the extent of this fallout and to assess its  impact on the environment, foodstuffs, and on the residents in the affected areas. In Fukushima Prefecture, various studies of external as well as internal exposures have been conducted since 2011~\cite{nagataki2013,hayano2013}. Particularly important in assessing the effect of radiation on the residents is to conduct personal dosimetry: One of the earliest reports was by Yoshida et al.~\cite{yoshidaimportance2012}, who measured the individual doses of the medical staff dispatched from Nagasaki to Fukushima City from March to July 2011.  They reported that the personal dose equivalent $H_{\rm P}$(10) ranged from 0.08 to 1.63 $\mu$Sv/h, significantly lower than the ambient dose equivalent rate  $H^*$(10) recorded by a monitoring station in Fukushima city which ranged from 0.86 to 12.34 $\mu$Sv/h.  

Large-scale individual dose monitorings have been conducted by most municipalities in Fukushima Prefecture since 2011. For example, Fukushima City started to distribute radio-photoluminescence glass dosimeters (Glass Badge\textsuperscript\textregistered) to school children and pregnant women in the fall of 2011, and the monitorings have been repeated every year. The percentage of the subjects whose {measured} ``additional'' dose was below 1~mSv/y was 51\% in 2011, 89\% in 2012, and 93\% in 2013. In 2014, 95.57\% of the 46,436 subjects were found to be below 1~mSv/y~\cite{fukushimagb}\footnote{The reduction in the external dose is due partly to the decay of $^{134}$Cs, having a half life of 2 years, and also to the decontamination efforts.}.

Such individual dose monitoring using passive dosimeters report a cumulative dose over a period of time, typically three months, to the participant; it is not possible to tell when and where the major contribution to the cumulative dose was received. In the present study, we therefore used active (solid-state) personal dosimeters  called ``D-shuttle'', which can record the integrated dose for each hour (hourly dose). The D-shuttles had already been used successfully in some studies. For example, Hayano et al.~\cite{hayanodshuttle} demonstrated the effectiveness of using D-shuttles to communicate the exposure situation to residents, and Naito et al.~\cite{naitoevaluation2014} used D-shuttles together with global-positioning system (GPS) receivers to compare individual versus ambient dose equivalent rates.

In the present study, 216 high-school students and teachers wore D-shuttles and kept journals of their behaviour for two weeks in 2014, and the external individual doses thus obtained were compared across the regions. This study was motivated and initiated by the high-school students living in Fukushima who wished to compare their own individual doses with those of people living in other parts of Japan, and also in other countries.

\section{Materials and Methods}

\begin{figure*}
\centering\includegraphics[width=\textwidth]{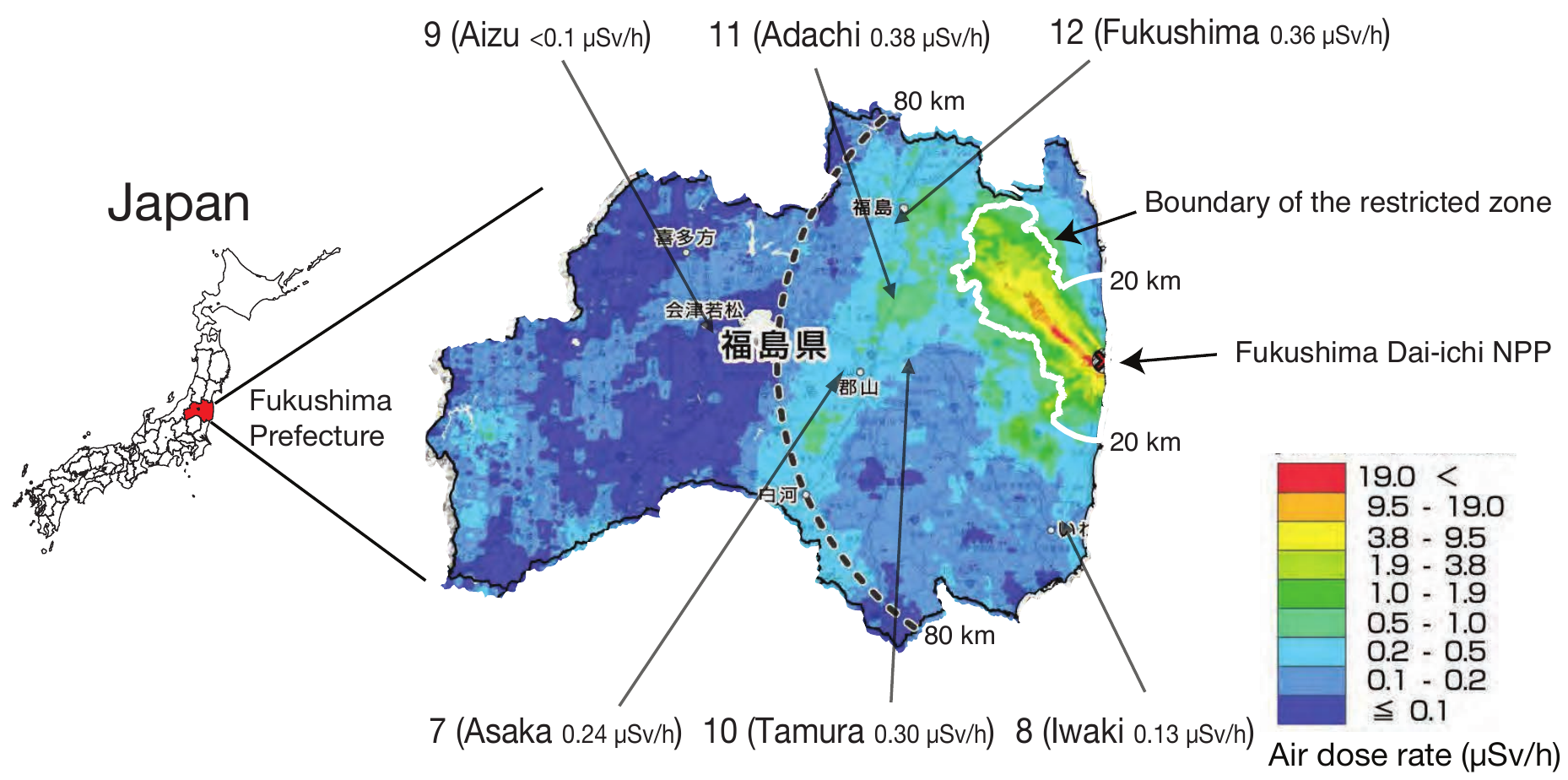}
\caption{\label{fig:fukushima} (left) The location of Fukushima Prefecture within Japan. (right) A map showing the air-dose rate ($\mu$Sv/h) at 1m above the ground estimated from the 9th airborne survey (as of November 7, 2014)~\cite{airborne}. The boundary of the evacuation zone is shown in white. The locations of the six high schools participated in the study are also shown { together with the air-dose rates estimated from the 9th airborne survey}.}
\end{figure*}

\begin{figure}
\centering\includegraphics[width=\columnwidth]{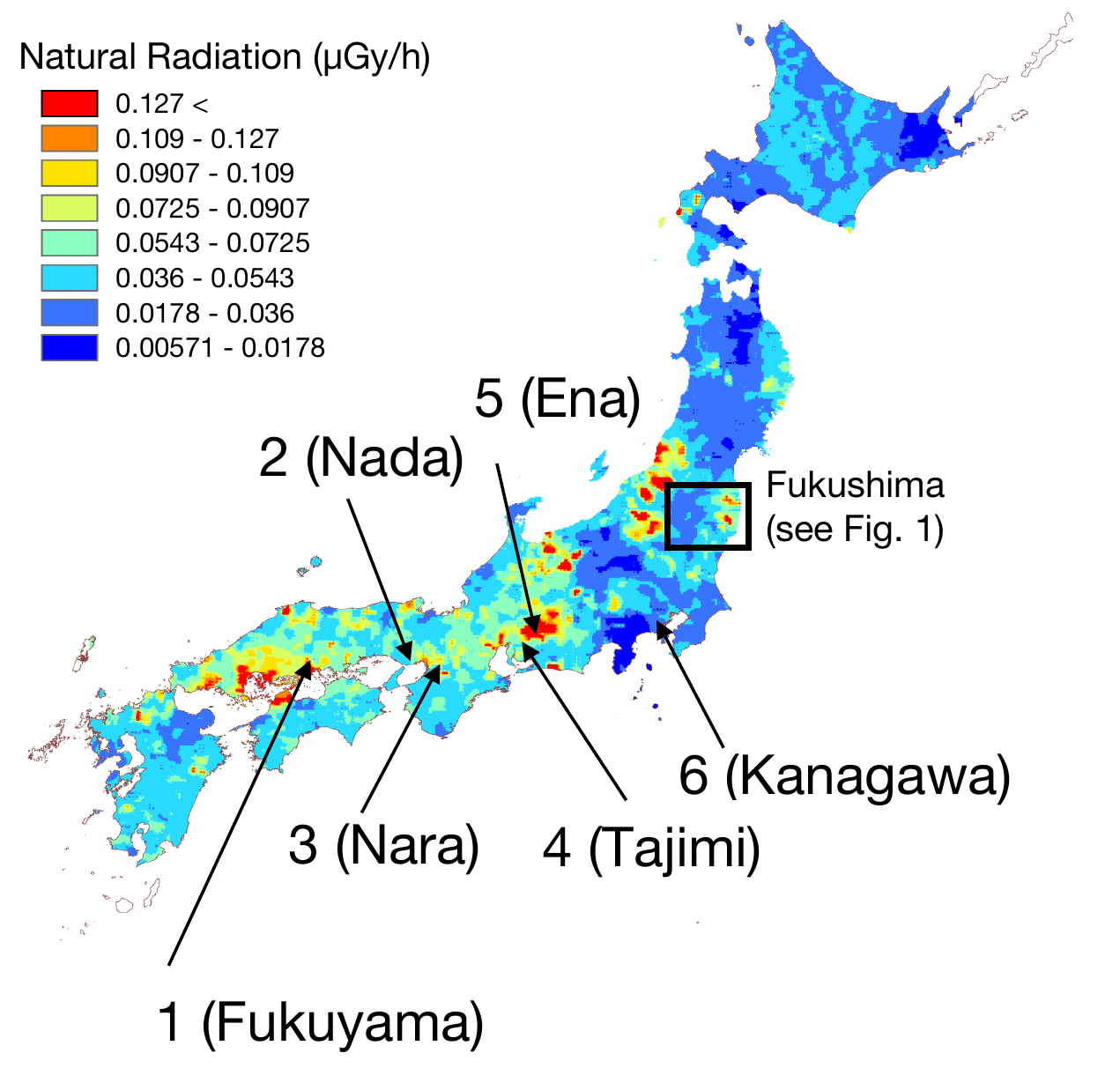}
\caption{\label{fig:outoffukushima} The natural radiation level map of Japan (in nGy/h) calculated from the chemical analyses of the soil samples by adding contributions from uranium, thorium and potassium-40~\cite{minatoterrestrial2010}. The map was adopted from Ref.~\cite{mapurl}. Note that the colour coding schemes are different between this figure and that in Fig~\ref{fig:fukushima}.  }
\end{figure}

\begin{figure}
\centering\includegraphics[width=\columnwidth]{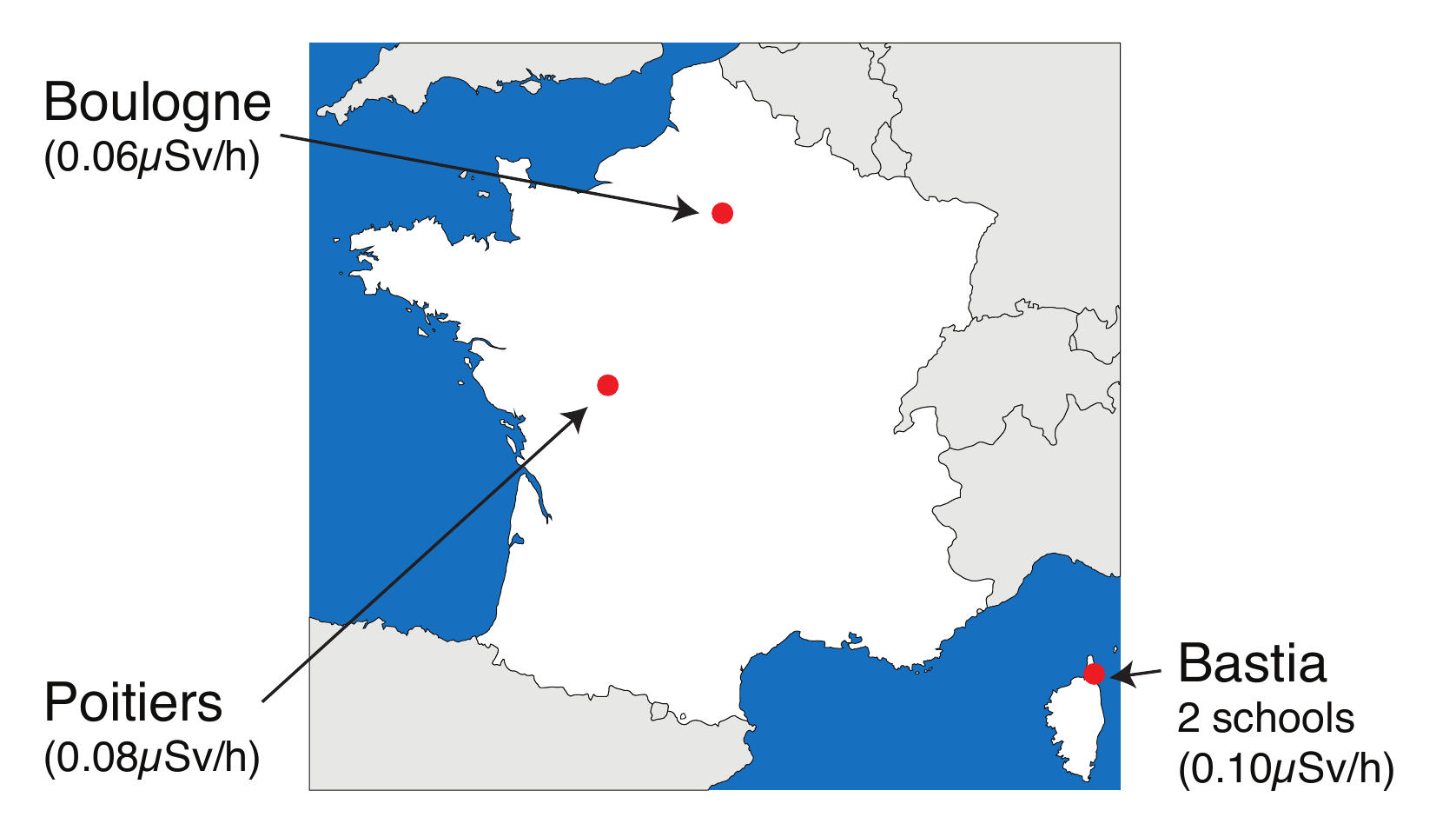}
\caption{\label{fig:france_background} The locations of the participating schools in France, and their nearby air dose rates (obtained from the IRSN ambient dose monitor)~\cite{irsnmonitor}}
\end{figure}

\begin{figure}
\centering\includegraphics[width=\columnwidth]{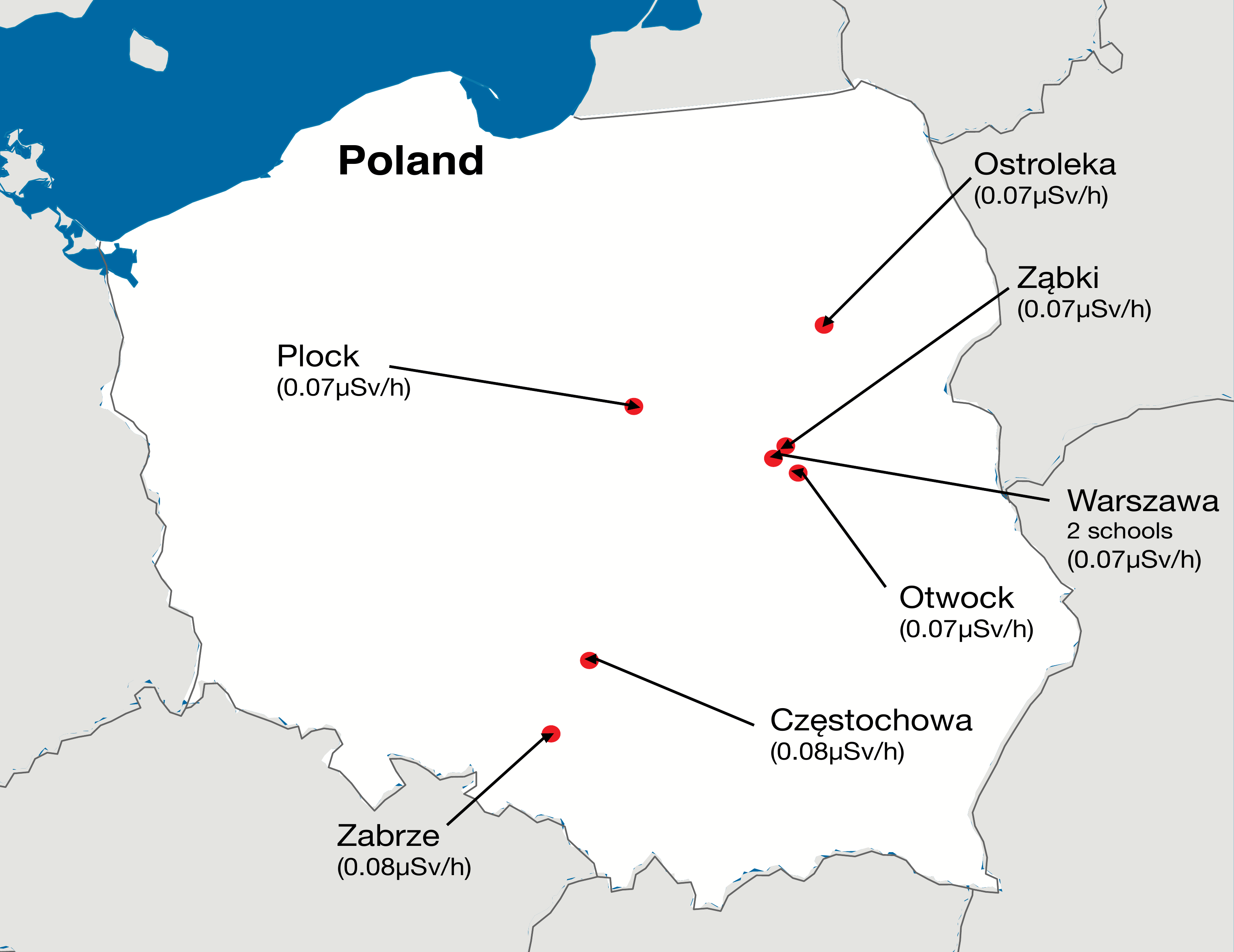}
\caption{\label{fig:poland_background} The locations of the participating schools in Poland, together with the average air dose rate of the county in which each school is located (obtained from Ref.~\cite{polandatlas})}
\end{figure}

\begin{figure}
\centering\includegraphics[width=\columnwidth]{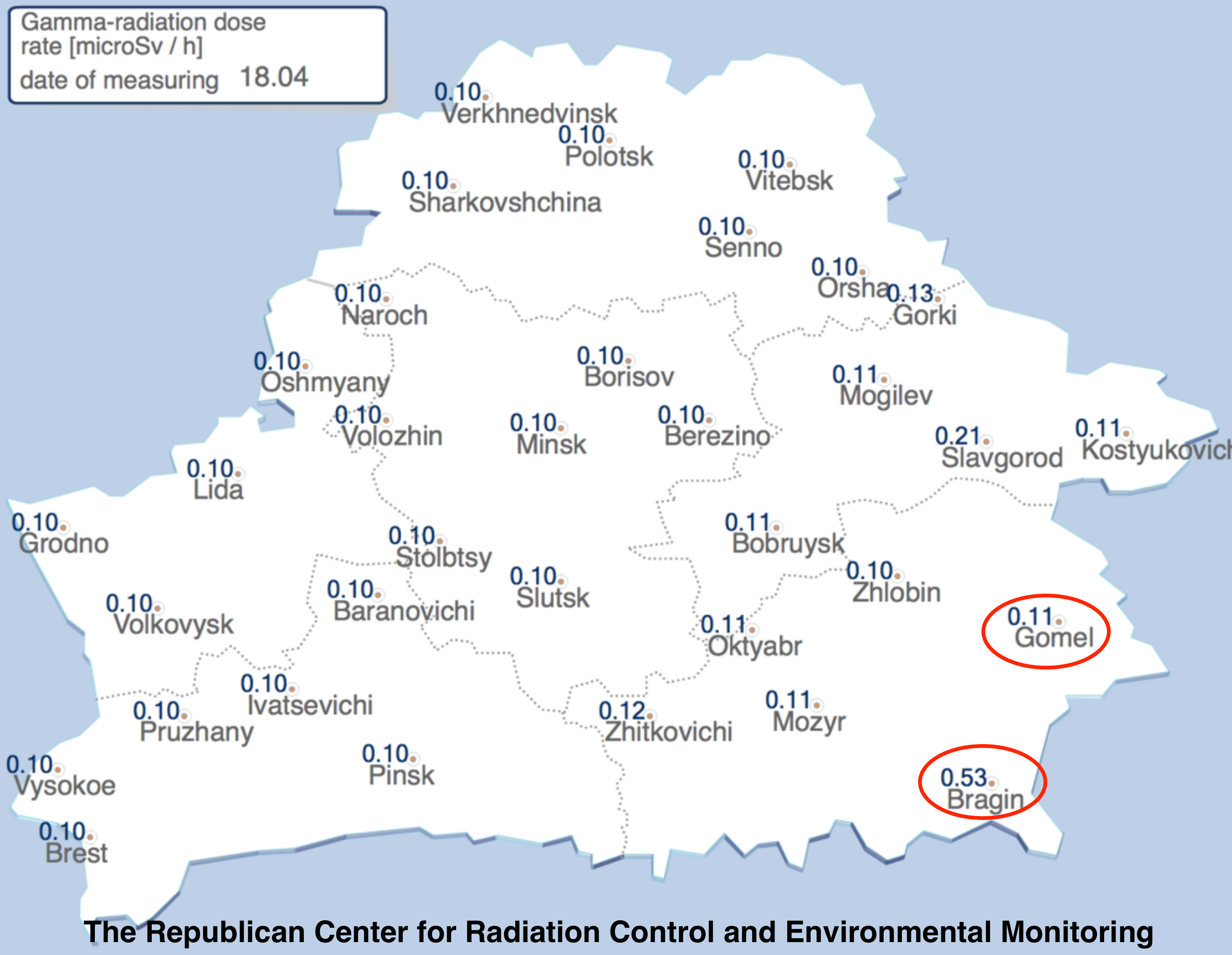}
\caption{\label{fig:belarus} The ambient dose equivalent rates in Belarus (April 2015), obtained from Ref.~\cite{belarusdose}, together with the locations of the participating schools.}
\end{figure}

The measurements were carried out by high-school students and teachers from twelve Japanese high schools (six in Fukushima Prefecture, see Fig.~\ref{fig:fukushima}, and six outside of Fukushima, see Fig.~\ref{fig:outoffukushima}), four high schools (three regions) in France (Fig.~\ref{fig:france_background}), eight high schools (seven regions) in Poland (Fig.~\ref{fig:poland_background}) and two high schools in Belarus (Fig.~\ref{fig:belarus}).
The total number of participants was 216, and the measurement period was two weeks { during the school term in each country.} 
.

The six Japanese schools outside Fukushima Prefecture were chosen by consulting the ``Geological Map of Japan'' (Fig.~\ref{fig:outoffukushima}, a natural radiation level map published by the geological society of Japan~\cite{mapurl,minatoterrestrial2010}). Fukuyama (labelled 1. in Fig.~\ref{fig:outoffukushima}), Tajimi (4.) and  Ena (5.) are in the region where the natural { terrestrial} background radiation level is relatively high, while Nada (2.), Nara (3.) and  Kanagawa (6.) are in the low-background region.

In Fukushima Prefecture, based on the airborne dose-rate monitoring map (Fig.~\ref{fig:fukushima}), schools in major cities, Fukushima, Nihonmatsu, Koriyama, Iwaki, Aizu were selected. Note that some 100,000 people were forced to evacuate from the restricted zone (indicated by the white border in Fig.~\ref{fig:fukushima}), who, after four years of the accident, cannot yet return to their homes. As such, the present study does {\bf not} include high schools in the restricted zones. When choosing the participants from Fukushima Prefecture high schools, care were taken so as to choose students living in various areas, house types (wooden vs concrete), and in their extracurricular activities.

In France and Poland, the high schools involved in the study participated on a voluntary basis without specific selection. In France, the four high schools are located in three different regions characterized by a range of natural {terrestrial} radiation background level, the lowest level being observed in Boulogne (closed to Paris) while higher value is observed in Corsica (see Figure~\ref{fig:france_background}). In Poland, the location of the schools is also ranging from lower values for high schools in the region of Warszawa {  to the highest in Zabrze} (see Figure~\ref{fig:poland_background}).

Two high schools from Belarus were involved due to their location in the Gomel region, impacted by the fallout of the Chernobyl accident. The first high school is located in Gomel City while the second one is located close to the exclusion zone (in Bragin district) and thus characterized by a higher ambient radiation dose rate (see Figure~\ref{fig:belarus}). 

\begin{figure}
\centering\includegraphics[width=0.8\columnwidth]{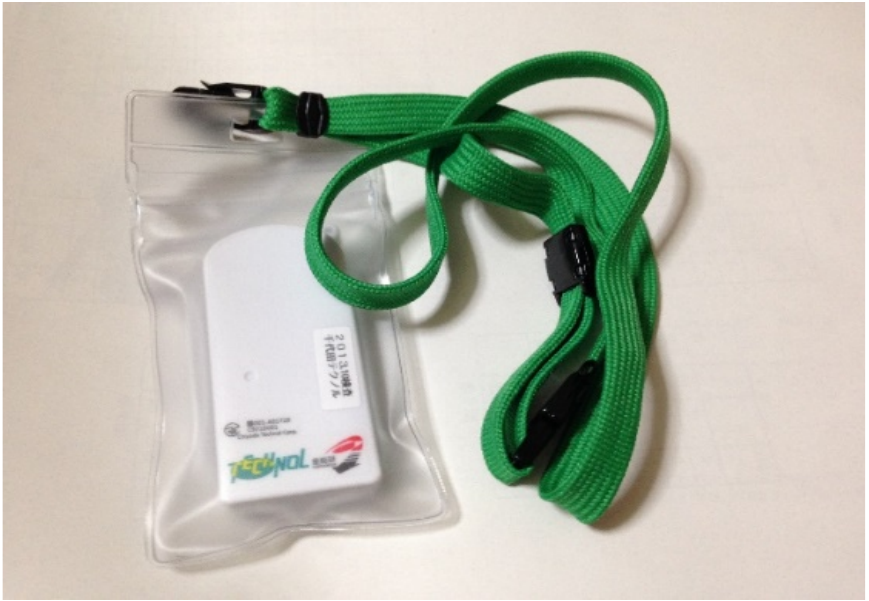}
\caption{\label{fig:dshuttle} A semiconductor-type personal dosimeter ``D-shuttle'', developed jointly by AIST and Chiyoda Technol Corporation. Each participant was instructed to wear the dosimeter on their chest, using the provided strap. 
}
\end{figure}

The individual dose-meter, called ``D-shuttle'' (FIg.~\ref{fig:dshuttle}),  developed jointly by the National Institute of Advanced Industrial Science and Technology (AIST) and Chiyoda Technol Corporation is 
a light (23g) and compact (68 mm (H) $\times$ 32 mm (W) $\times$ 14 mm (D)) device, 
based on a $2.7 \times 2.7$ mm$^2$ silicon sensor, and is capable of logging the integrated dose every hour in an internal memory with time stamps~\cite{dshuttle}. The memory can be later read out by using a computer interface. By comparing the data with the activity journal kept by the participant, we can analyse the relationship between the personal dose and behaviour (when, where, and what) of the participant.  Each D-shuttle was calibrated with a $^{137}$Cs calibration source for $H_{\rm P}$(10),
In accordance with the International Organization for Standardization (ISO 4037-3)~\cite{iso4037}.
The relative response is ±30\% from 60 keV to 1.25 MeV, and its least detectable value was  0.01$\mu$Sv/h. The $H_{\rm P}$(10) measured with a personal dosimeter such as D-shuttle is known not to underestimate the effective dose $E$ (i.e.,  $H_{\rm P}(10) \gtrsim E$)   even in cases of lateral or isotropic radiation incidence on the body, as in the present study~\cite{icrp103}.

Although the D-shuttle is well shielded against external electromagnetic noise and is protected against  mechanical vibration, occasional spurious ``hits'' are unavoidable. These typically show up as an isolated large ``spike'' in the readout data. In such cases, we checked the activity journal and queried the participant to determine whether or not the ``spike'' was likely caused by radiation, as will be discussed in detail in section~\ref{sec:outliers}.

Participants were instructed to always wear the D-shuttle on his/her chest, except during sleep when the unit was left near the bedside. In Table~\ref{tab:rawdata}, typical data read out from the D-shuttle are shown together with a part of the activity journal.
The number of data points per participant was 24 h $\times 14$ d = 336. As there were some participants who could not take part in the measurements 
for the full  14 days, the total number of data points for the  216 people were 70,879.

The D-shuttle records both natural radiation and the radiation due to the accident ($^{134,137}$Cs); { the latter contribution is negligible} except in Fukushima Prefecture and in Belarus. 
When comparing the individual doses across the regions, we used the recorded doses by the D-shuttle including both doses from natural background radiation { and radiation from radiocaesiums\footnote{This is unlike the glass-badge measurements conducted in Fukushima, which report ``additional'' dose after { terrestrial} background subtraction.}, since these two contribute inseparably and additively to the individual doses}. In this way, in addition to using the same device and standardising the measurement protocol, it is possible to compare the individual doses across all the participating regions.

\begin{table}
\begin{small}
\caption{\label{tab:rawdata} A typical example of a D-shuttle data.}
\begin{tabular}{cc|c}
Date and time& Hourly dose ($\mu$Sv)& Location\\
\hline
2014/06/27 15h &0.12&school\\
2014/06/27 16h & 0.07&school\\
2014/06/27 17h&0.10&school\\
2014/06/27 18h&0.10&school\\
2014/06/27  19h&0.14&school\\
2014/06/27 20h&0.04&home\\
2014/06/27 21h &0.06&home\\
2014/06/27 22h&0.12&home\\
2014/06/27 23h&0.13&home\\
2014/06/28 00h&0.07&home\\
\hline
\end{tabular}
\end{small}
\end{table}

\section{Results}

\subsection{Comparison of the hourly dose across regions}
\begin{figure}[bht]
\centering\includegraphics[width=\columnwidth]{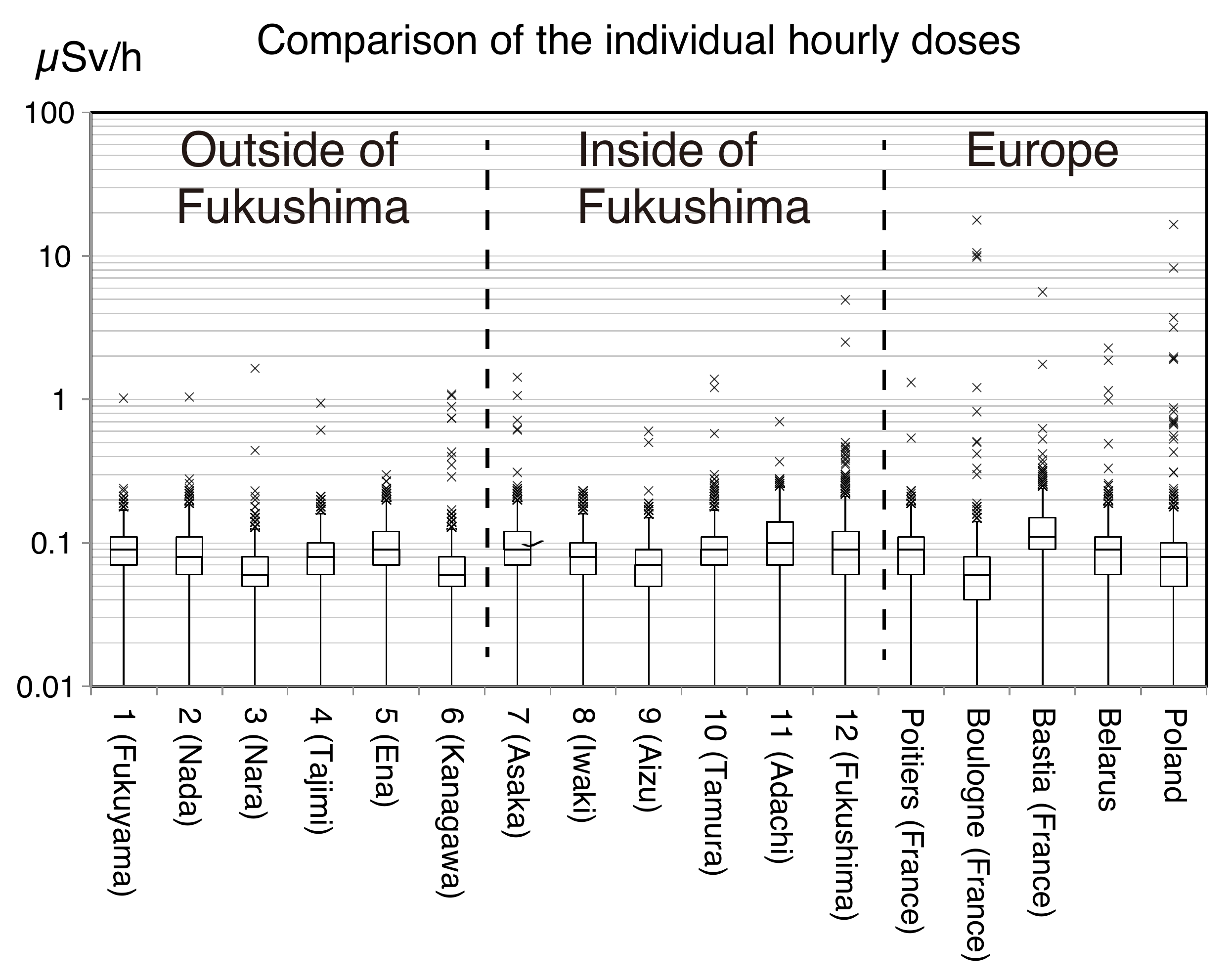}
\caption{\label{fig:comparison1} Box-and-whisker diagrams of the individual hourly dose ($\mu$Sv/h) for 6 schools outside Fukushima, 6 schools in Fukushima, and five European regions. The ordinate is in logarithmic scale. For the definitions of the boxes, whiskers and outliers ($\times$), see the main text.
}
\end{figure}

For each school or region, the number of participants ranged from 10-33, and the number of data points were from $\sim 3300 - 9800$, as summarised in Table~\ref{tab:dosetable}.
Fig.~\ref{fig:comparison1} shows the individual hourly dose ($\mu$Sv/h) distributions for 12 Japanese schools and 5 European regions in the form of a box-and-whisker plot. Note that the vertical axis is in logarithmic scale. The bottom of the box represents the first quartile, top of the box represents the third quantile and the centre line represents the median. The length of the whisker is 1.5 times the height of the box (when the lower end of the whisker falls below 0.01$\mu$Sv/h it is set to 0.01$\mu$Sv/h). The outliers (data points larger than the top end of the upper whisker) are indicated by crosses. The percentage of outliers was 1.5\% { and the percentage of outliers exceeding $1 \mu$Sv/h was 0.045\%.  Most of the extreme outliers ($\geq 1 \mu$Sv/h) are not accident related, as they are found in all the regions, and, as will be discussed in Sec.~\ref{sec:outliers}, they are due to noise. We did not however exclude them in subsequent analyses, as i) it was not possible, by consulting the journals and sometimes conducting interviews, to determine what caused the outlier in every case, and, ii) including them did not change the median annualised dose values (Sec.~\ref{sec:annualised}).} The numerical values of the first quartile, the median, and the third quartile are provided in Table~\ref{tab:dosetable}.

\begin{table*}
\begin{footnotesize}
\caption{\label{tab:dosetable} The number of participants, the number of data points, and individual doses (25 percentile, median and 75 percentile) for each school/region.}
\begin{tabular}{l|c|r|r|rrr}
\hline
&&&\multicolumn{3}{|c}{Individual dose ($\mu$Sv/h)}\\
 School or region&Period&No.\ &No.\ of&25 percentile&Median&75 percentile\\
 &&participants&data points &&&\\
\hline
\hline
Fukuyama	&2014.06.18$\sim$07.01&11&3696&0.07&0.09&0.11\\
Nada&2014.06.18$\sim$07.01&11&3696&0.06&0.08&0.11\\
Nara&2014.06.18$\sim$07.01&10&3360&0.05&0.06&0.08\\
Tajimi&2014.06.18$\sim$07.01&10&3360&0.06&0.08&0.10\\
Ena&2014.06.18$\sim$07.01&10&3360&0.07&0.09&0.12\\
Kanagawa&2014.06.18$\sim$07.01&11&3696&0.05&0.06&0.08\\
Asaka&2014.06.18$\sim$07.01&10&3696&0.07&0.09&0.12\\
Iwaki&2014.06.18$\sim$07.01&11&3696&0.06&0.08&0.10\\
Aizu&2014.06.18$\sim$07.01&11&3696&0.05&0.07&0.09\\
Tamura&2014.06.18$\sim$07.01&11&3696&0.07&0.09&0.11\\
Adachi&2014.06.18$\sim$07.01&11&3696&0.07&0.10&0.14\\
Fukushima&2014.06.18$\sim$07.01&14&4704&0.06&0.09&0.12\\
\hline
Boulogne&2014.11.06$\sim$11.18&11&3278&0.04&0.06&0.08\\
Poitiers&2014.11.06$\sim$11.19&16&5168&0.06&0.09&0.11\\
Bastia&2wks between 2014.11.05$\sim$11.20&13&4276&0.09&0.11&0.15\\
Belarus&2014.10.07$\sim$10.20 &12&4032&0.06&0.09&0.11\\
& (4 were between 10.15$\sim$10.28)&&&&\\
Poland&2 wks between 2014.11.17$\sim$12.12&33&9773&0.05&0.08&0.10\\
\hline
\hline
\end{tabular}
\end{footnotesize}
\end{table*}

Median hourly individual doses for participants from six high schools in Fukushima Prefecture are $0.07-0.10 \mu$Sv/h, while those for participants from outside Fukushima Prefecture are $0.06-0.09\mu$Sv/h. The median hourly individual doses for participants from France, Poland and Belarus are $0.06-0.11\mu$Sv/h. 

Within Japan, the hourly dose distribution for Asaka in Fukushima and that for Ena (outside of Fukushima) are almost the same. The hourly dose distribution for Aizu in Fukushima is close to that of the low-dose regions outside of Fukushima, e.g., Nara and Kanagawa. { These show that the hourly individual dose distributions of these regions in Fukushima Prefecture are not significantly higher than in those of other parts of Japan.}

Within Europe, the hourly dose distribution of Bastia (Corsica, France) is similar to or even slightly higher than those in Fukushima Prefecture. This is consistent with the known fact that Bastia is in a region where the natural radiation dose is relatively high. 

\subsection{Annualised comparison of individual dose values\label{sec:annualised}}
\begin{figure}[bht]
\centering\includegraphics[width=\columnwidth]{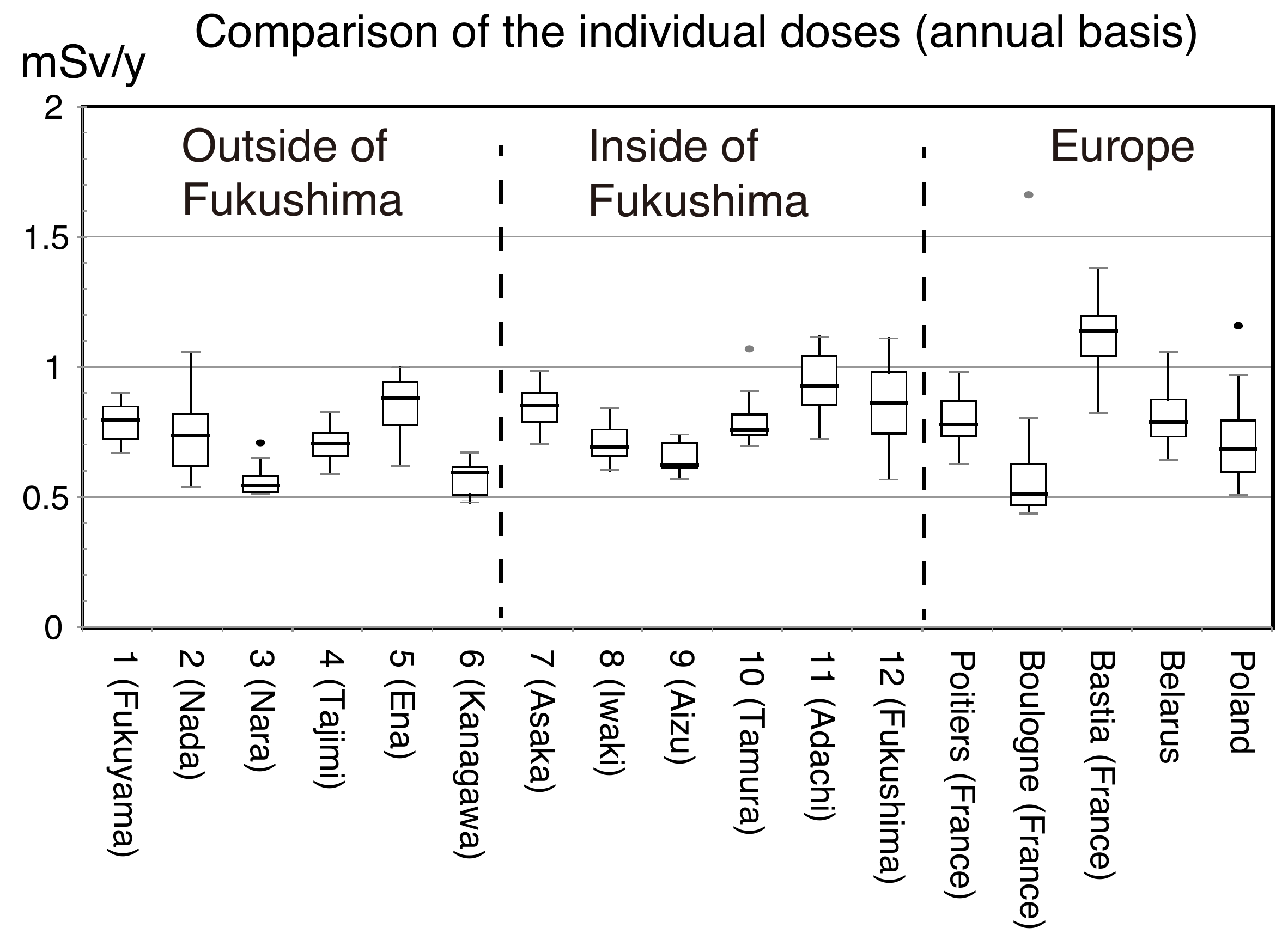}
\caption{\label{fig:comparison3} The two-week integrated individual dose was converted to annual dose (mSv/y), presented in box-and-whisker diagrams. For the definitions of the boxes and whiskers, see the main text.
}
\end{figure}

For each participant, we integrated the individual hourly dose over the two-week measurement period, which was then multiplied by a factor 365/14 to estimate the annual dose. In the integration, we did not exclude outliers. Fig.~\ref{fig:comparison3} compares the distributions of estimated annual doses in box-and-whisker plots for the 12 Japanese schools and 5 European regions. As before, the bottom (top) of the box represents the first (third) quantile, and the horizontal line represents the median. The upper ends (bottom ends) of the whiskers indicate the maximum (minimum) values.

\begin{figure}
\centering\includegraphics[width=\columnwidth]{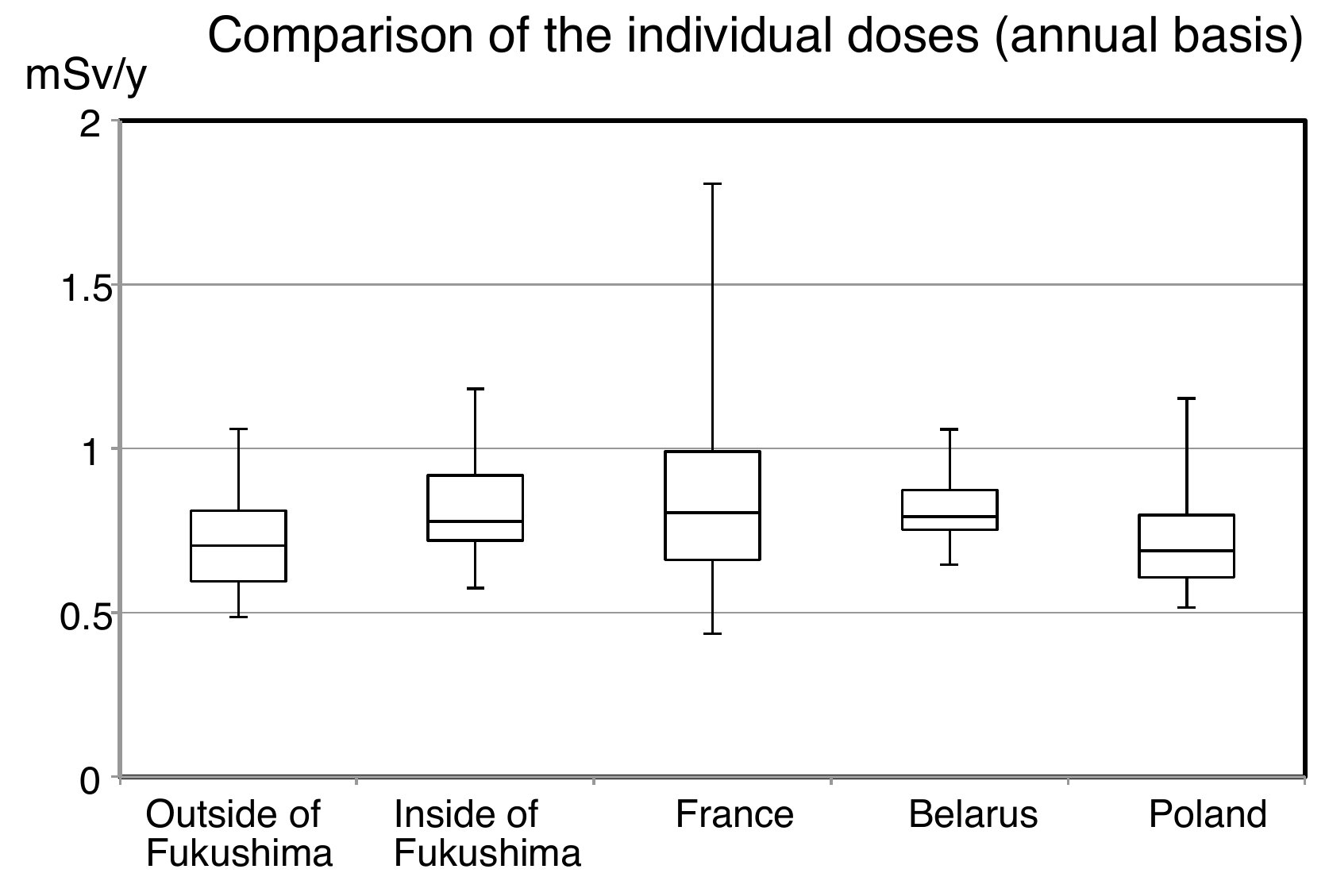}
\caption{\label{fig:comparison4}  Same as Fig. \ref{fig:comparison3}, but is now grouped to show 1. outside of Fukushima, 2. in Fukushima, 3. France, 4. Belarus and 5. Poland.}
\end{figure}

The median of the estimated annual doses in Fukushima are $0.63 - 0.97$mSv/y, those in other prefectures are $0.55 - 0.87$mSv/y and those  in France, Poland and Belarus are $0.51 - 1.10$mSv/y.  
Note again that all these comparisons were done including natural radiation.  Fig.~\ref{fig:comparison4}, shows that the annual individual doses of high-school students in Fukushima are not much higher than those in other regions, but are similar to the natural radiation level in other parts of the world and in the same range of those observed in Gomel region 28 years after the Chernobyl accident.

\subsection{Analysis of the ``outliers''\label{sec:outliers}}

\begin{figure}
\includegraphics[width=0.35\columnwidth]{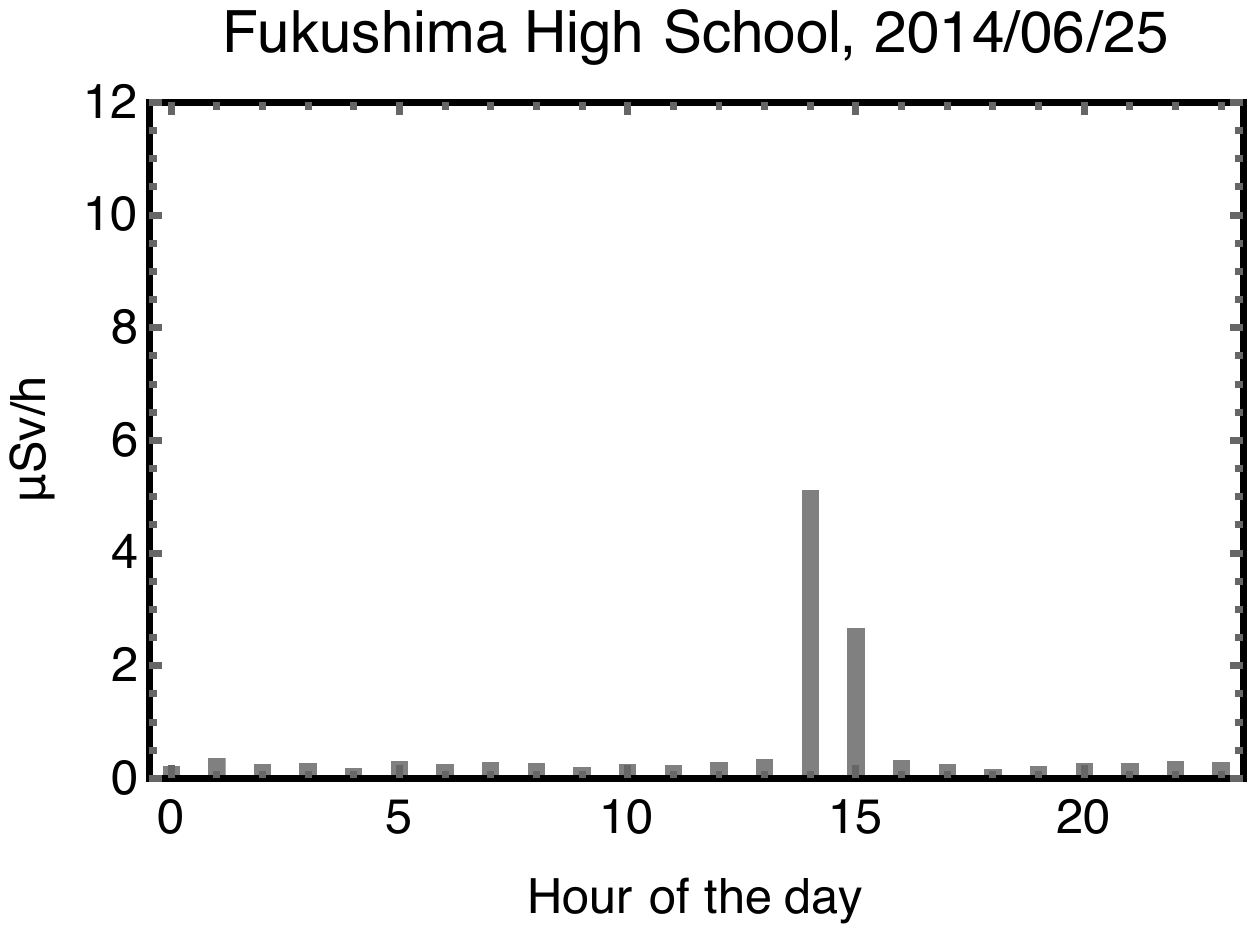}
\vspace*{5mm}

\includegraphics[width=0.35\columnwidth]{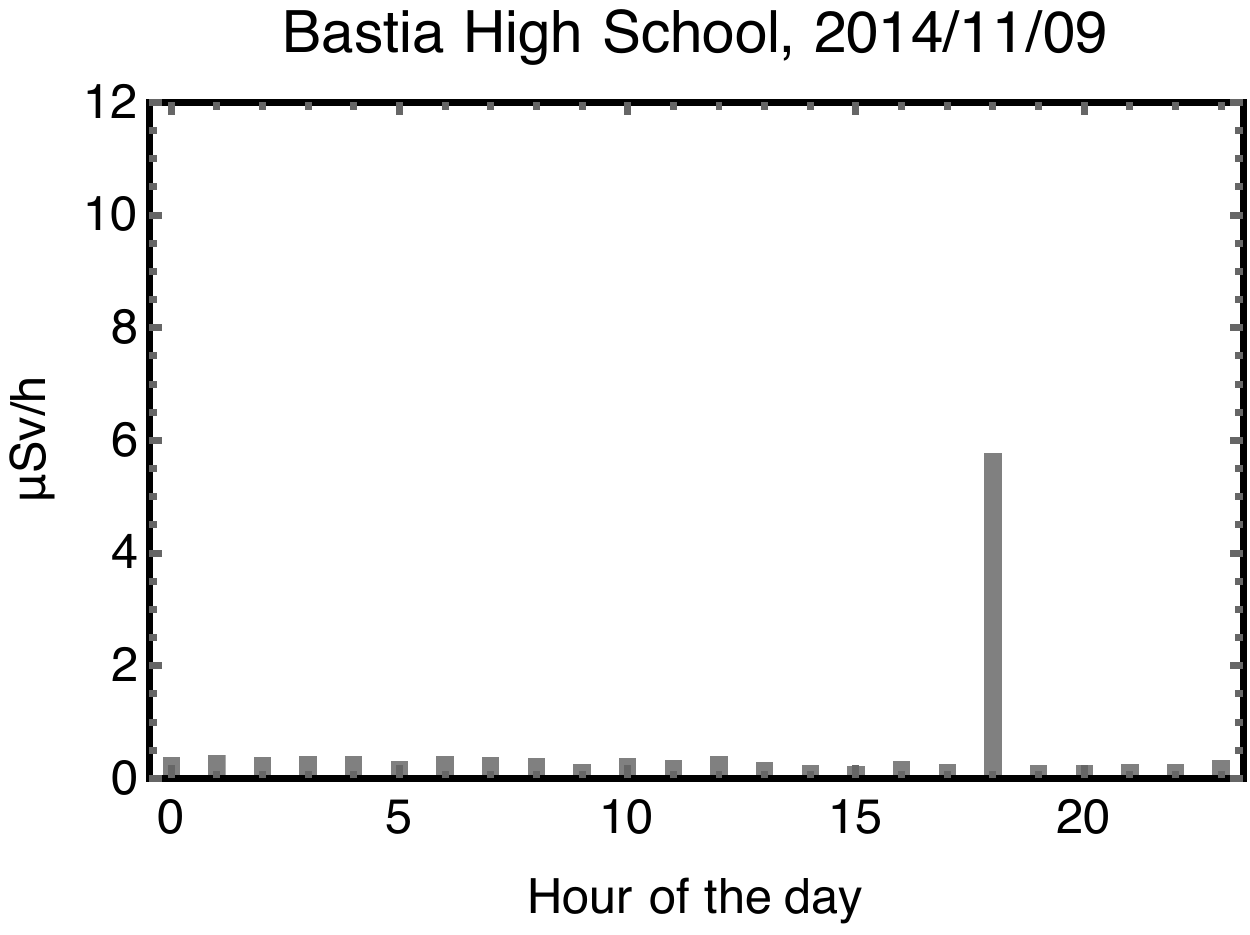}
\vspace*{5mm}

\includegraphics[width=0.35\columnwidth]{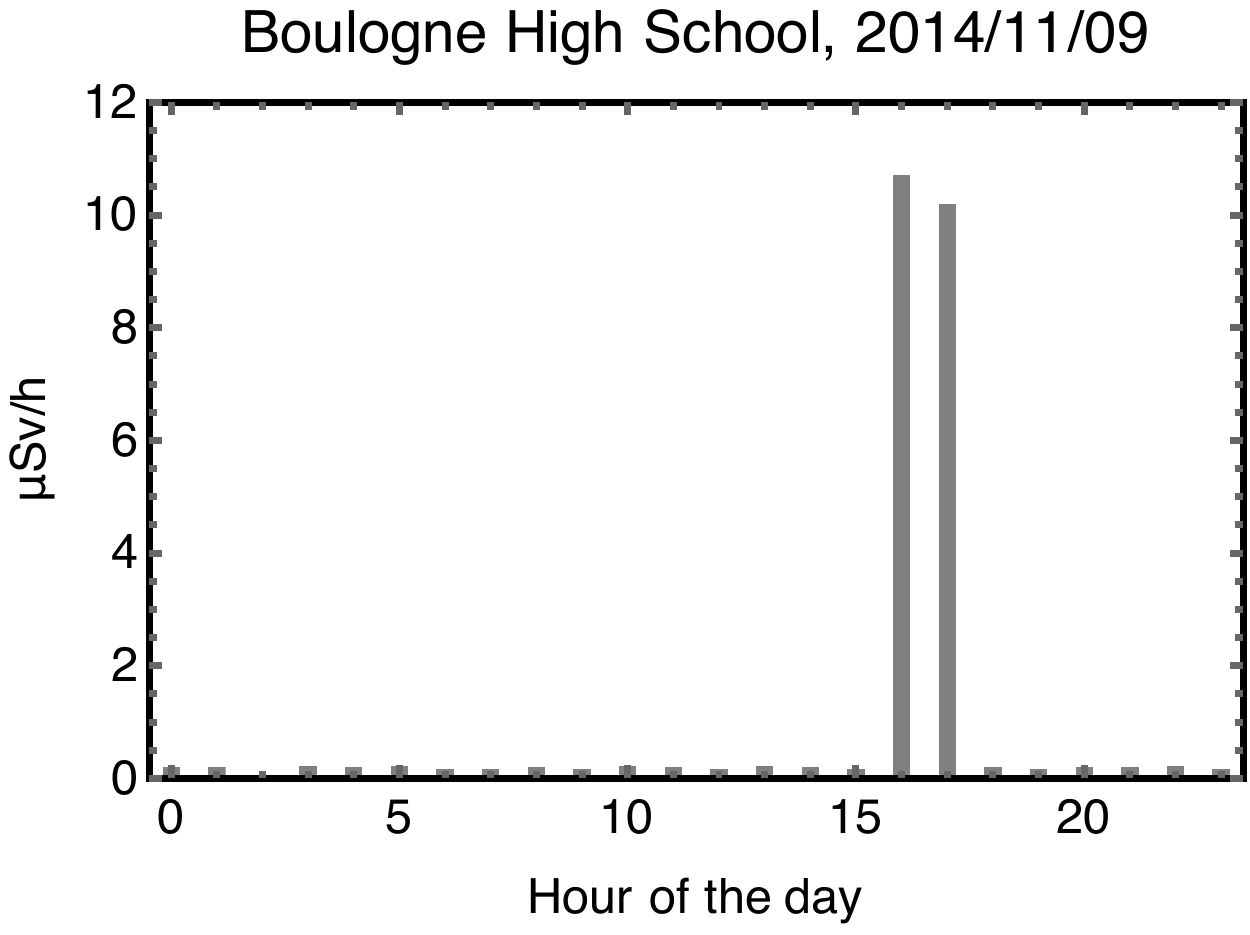}
\caption{\label{fig:noise2} (top) Time variation of the hourly dose for a participant from Fukushima high school. (middle) Time variation of the hourly dose for a participant from Bastia high school. (bottom) Time variation of the hourly dose for a participant from Boulogne high school.}
\end{figure}

We here discuss what may have caused the ``outliers'' found in Fig.~\ref{fig:comparison1}. { Some are due to high radiation level, and others are due to noise. In Fig.~\ref{fig:noise2}, we show three typical examples from Fukushima (top), Bastia (middle) and Boulogne (bottom).}

A large value of 5$\mu$Sv/h was recorded for one participant from Fukushima high school (Fig.~\ref{fig:noise2} (top)). This was when this person (teacher) visited Okuma town in the restricted zone,  close to the Fukushima Dai-ichi Nuclear Power Plant (Fig.~\ref{fig:fukushima}) for research purposes. 
For two hours, 15:00 and 16:00, high hourly doses were recorded, and this coincided with the activity journal entry of this person.

{ The large values found in the data of participants from Bastia and Boulogne are due to noise, since interviews revealed no reasons for high radiation exposures. As was discussed above, occasional malfunction of the D-shuttle device due to  electromagnetic/mechanical noise is unavoidable, which usually shows up as a single isolated event as shown in Fig.~\ref{fig:noise2} (middle). 
Large values were recorded for two consecutive hours, however, in the case of a participant from Boulogne (Fig.~\ref{fig:noise2} (bottom), 16:00-17:00).   Interview of this person revealed nothing notable, and we concluded that this reading also must have been noise, although having noise for two consecutive data points is quite rare (this is the only such case in $>70,000$ data points).
}

\subsection{Analysis of hourly doses versus behaviour}

\begin{figure}
\centering\includegraphics[width=0.35\columnwidth]{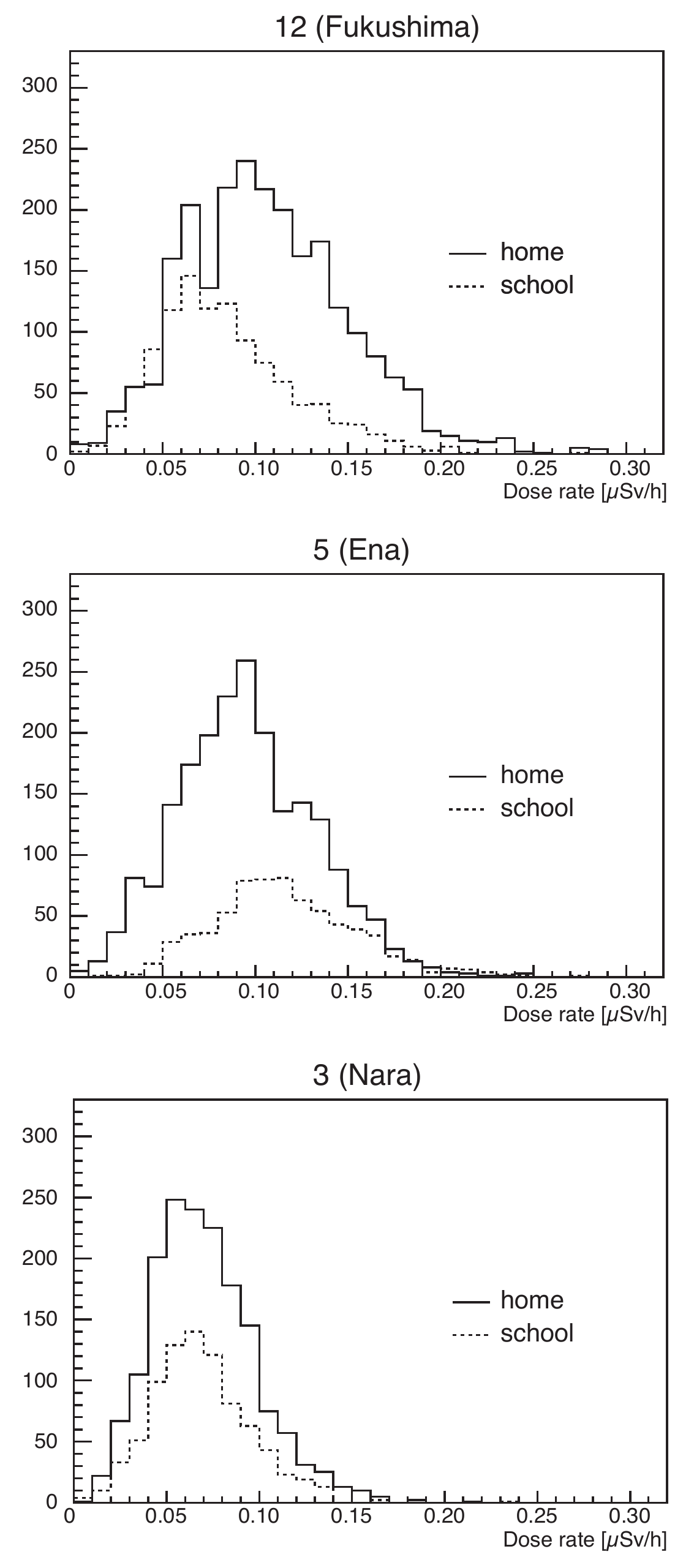}
\caption{\label{fig:inout} The activity journal entries were used to extract the dose-rate data at home and in the school building, and the dose-rate histograms are plotted separately (home: solid, school: dotted). The ordinate is the frequency and abscissa is the hourly dose ($\mu$Sv/h). }
\end{figure}

Fig.~\ref{fig:inout}
compares the hourly dose distributions for three schools, Fukushima, Ena and Nara. 
The activity journal was used to extract the dose-rate data at school (home), and they are shown in the solid (dotted) histograms. The ordinate is the frequency, and the abscissa is the hourly dose ($\mu$Sv/h). 

For students attending Fukushima high school, the hourly doses were lower at school and higher at home (also see Table~\ref{tab:inout}). This may be due to the fact that most of the homes of the participants are wooden, while the school building is made of concrete and surrounding school ground was decontaminated. On the other hand, in Ena, the hourly doses were higher at school rather than at home.  This was found to be due to the relatively high concentration of natural radioactivity in the   building materials of the school (granite). In Nara, there were no noticeable differences between the two distributions, and both were lower than in the other two schools.  It has to be noticed that for the schools from Europe, no significant variation is observed according school building or home. The only variation is due to the time spent outside building according the region characterized by the {terrestrial} background radiation level. 

\begin{table}

\caption{\label{tab:inout}The mean hourly dose at school (indoor) and home (indoor) for students attending Fukushima, Ena and Nara high schools.\\}
\begin{tabular}{lrr}
School&School ($\mu$Sv/h) & Home  ($\mu$Sv/h)\\
\hline
Fukushima &  0.080&0.102\\
Ena &0.111&0.091\\
Nara &0.064&0.066\\
\hline
\end{tabular}
\end{table}

\section{DISCUSSION}

\Cref{fig:comparison1,fig:comparison3,fig:comparison4} show that the personal doses for the high school students from Fukushima Prefecture were not significantly higher than in other regions and countries. This may seem surprising since there are still contributions from radioactive caesium in Fukushima, as airborne monitoring and other measurements clearly show. 

\begin{figure}
\centering\includegraphics[width=.8\columnwidth]{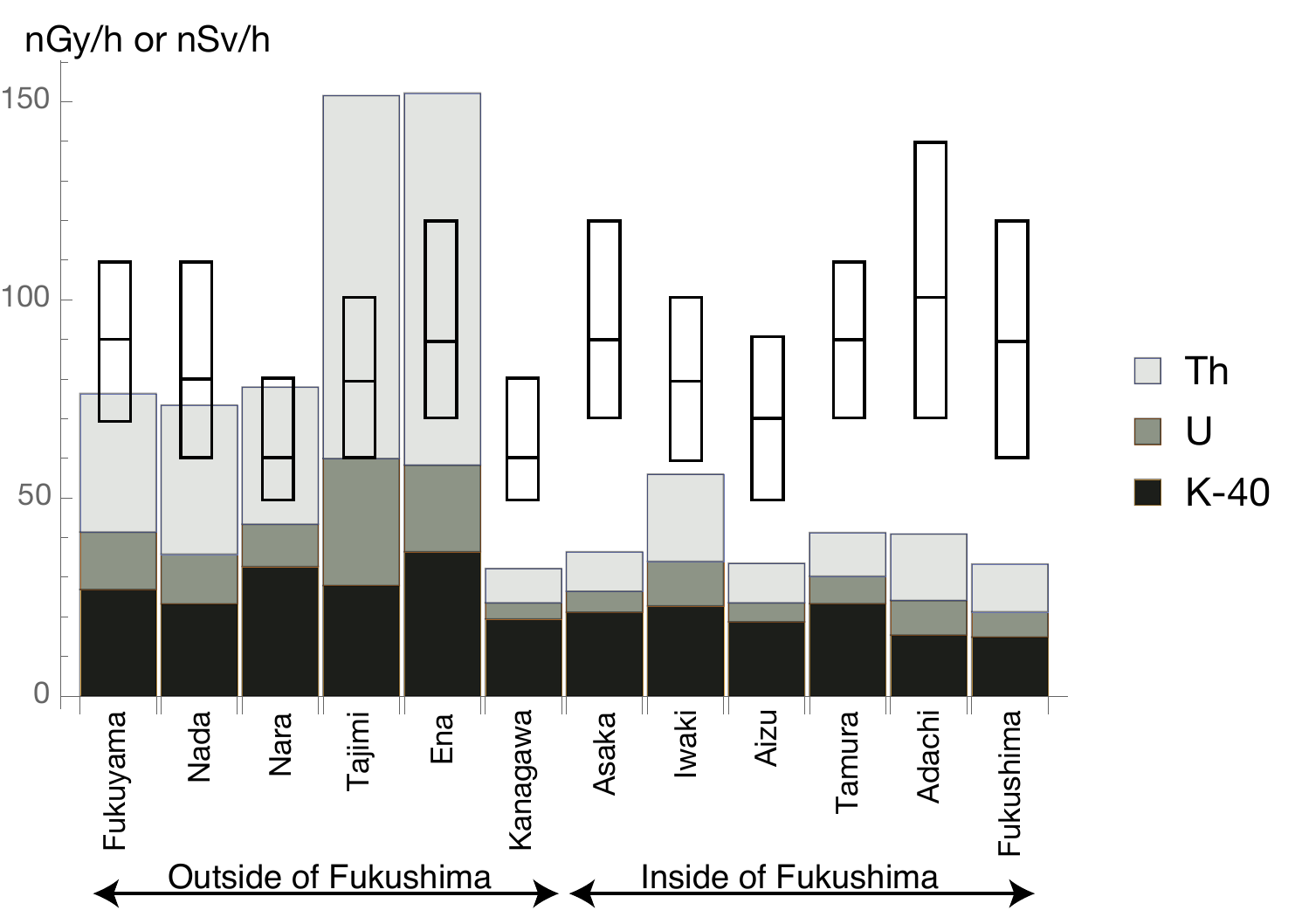}
\caption{\label{fig:estimateddose} The stacked bars show the estimated natural radiation level (nGy/h) around the participating schools, based on the soil-sample chemical analysis database {(N.B.\ since those soil samples were collected from riverbeds, the estimated radiation levels may not necessarily coincide with the typical values in residential areas})~\cite{soildatabase}. The individual hourly dose (nSv/h) distribution measured at each school is indicated by the box diagrams (same as in Fig.~\ref{fig:comparison1}).}
\end{figure}

In order to better understand the situation, we show in Fig.~\ref{fig:estimateddose} the estimated air kerma for the 12 participating schools in Japan, using the database of chemical analysis of soil samples~\cite{soildatabase}. We selected the soil-sample data within 5~km of participating schools from the database (when there were multiple sampling points, we took their average), and used the equation  $K=13.0 C_K+5.4 C_U+2.7C_{Th}$ to estimate the air kerma $K$ (nGy/h), with $C_K$ being the $^{40}$K concentration (in \%), $C_U$ being the uranium concentration (in ppm) and $C_{Th}$ being the thorium concentration (in ppm)~\cite{beck}. 
{ The relation between the air kerma and the effective dose varies depending on the irradiation angle to the body (c.f., ICRP Publication 74 (1996)\cite{icrp74}, figure 9). As people almost evenly receive terrestrial background radiation from all sides, rotational irradiation geometry is adequate for the relation. Although effective dose per unit air kerma (Sv/Gy) at rotational geometry is around 0.9 in the energy region of terrestrial background radiation (c.f., ibid. two-dot chain line of figure 8), it can be treated as unity without losing the rationality of our argument, since the difference lies within the uncertainty of the measurement.}

As shown, outside of Fukushima, the estimated effective dose rates from the terrestrial radiation air kerma rates and measured individual dose rates are correlated and have similar values. However, in Fukushima, the individual dose rates are higher than the estimated effective dose rates of terrestrial radiation. In fact, the terrestrial radiation background is low in Fukushima; the radiation due to the distributed radio-caesium was added on top of the terrestrial radiation, but that increment is  not high as might be expected. Thus, although the dose rate in most of Fukushima Prefecture was elevated by the nuclear accident, the total external individual dose rates observed for the Fukushima high school students involved in this study are not significantly different from those in other regions.

The natural radiation levels vary from region to region. In Japan, the nation-wide average of the terrestrial gamma-ray contribution to the effective dose is evaluated to be 0.33 mSv/y~\cite{why033}, lower than the world average of 0.48~mSv/y~\cite{unscear2000}.
In France, the average value is 0.47 mSv/y, similar to the world average but with variation from a factor 5 according the regions, ranging from about 0.2 to 1 mSv/y~\cite{frenchdose}.   In the present study, the D-Shuttle measured the sum of the natural radiation dose and the additional dose due to the nuclear accident, if any was detectable. Nevertheless, in Fukushima as well as in Belarus, the individual annual dose {\em including natural radiation} was below 1 mSv/y for most of the participating high-school students. It is interesting to mention that ICRP stated in Publication 111 that ``Past experience has demonstrated that a typical value used for constraining the optimisation process in long-term post-accident situations is 1 mSv/year''~\cite{ICRP111}.

\section{CONCLUSION}

Twelve high schools in Japan (of which six are in Fukushima Prefecture), four in France, eight in Poland and two in Belarus cooperated in the measurement and comparison of individual external doses in 2014.  In total 216 high-school students and teachers participated in the study. Each participant wore an electronic personal dosimeter ``D-shuttle'' for two weeks, and kept a journal of his/her whereabouts and activities. The median annual doses were estimated to be 0.63-0.97 mSv/y in Fukushima Prefecture, 0.55-0.87 mSv/y outside of Fukushima in Japan, 0.51-1.10 mSv/y in Europe (0.09 in Belarus), thus demonstrating that the individual external doses currently received by participants in Fukushima and Belarus are well within the {terrestrial} background radiation levels of other regions/countries. { The present study also demonstrated that the measurement of individual dose rates together with the use of activity journals is a powerful tool for understanding the causes of external exposures, and can be useful and clearly understandable tool for  risk communication for people living in  contaminated areas.}

\section{Author contributions}
This study was conceived by students of Fukushima high school, and was designed by { Hara T., Niwa O., Miyazaki M., Tada J., Schneider T., Charron S., and Hayano R. Data were analysed by  Onodera H., Kiya M., Suzuki K., Suzuki R., and  Saito M. Onodera H., Suzuki R., Saito M., Anzai S., and Fujiwara Y.\ wrote the manuscript in Japanese, and Hayano R., Niwa O., Schneider T., and  Tada J.\ finalised the English version. 
Technical details regarding the D-shuttle were provided by Ohguchi H}. All the students and teachers in the author list participated in the measurement, and contributed comments on the data analysis and the manuscript. This work was in part supported by the Super Science High School (SSH) programme of Japan Science and Technology Agency (JST).


\begin{thebibliography}{99}
\bibitem{tanakaaccident2012} S. TANAKA, ``Accident at the Fukushima dai-ichi nuclear power stations of TEPCO —Outline \& lessons learned—'', Proceedings of the Japan Academy, Series B 88, 471-484 (2012).

\bibitem{nagataki2013} S. Nagataki, N. Takamura, K. Kamiya, and M. Akashi, ``Measurements of Individual Radiation Doses in Residents Living Around the Fukushima Nuclear Power Plant'',  RADIATION RESEARCH 180, 439-447 (2013).

\bibitem{hayano2013} R.S. Hayano, M. Tsubokura, M. Miyazaki, H. Satou, K. Sato, S. Masaki, and Y. Sakuma, “Internal radiocesium contamination of adults and children in fukushima 7 to 20 months after the Fukushima NPP accident as measured by extensive whole-body- counter surveys'', Proceedings of the Japan Academy, Series B 89, 157-163 (2013).

\bibitem{yoshidaimportance2012} Kohji Yoshida, Kanami Hashiguchi, Yasuyuki Taira, Naoki Matsuda, Shunichi Yamashita, and Noboru Takamura, ``Importance of personal dose equivalent evaluation in fukushima in overcoming social panic'', Radiation protection dosimetry 151 (2012), 10.1093/rpd/ncr466.

\bibitem{fukushimagb} Result of the Fukushima-city glass badge survey, available in Japanese at \url{http://www.city.fukushima.fukushima.jp/uploaded/attachment/39637.pdf}, last accessed May 10, 2015.

\bibitem{hayanodshuttle} Hayano R S and Miyazaki M 2014 Internal and external doses in Fukushima: measuring and communicating personal doses FBNews No.447. 1–5 (in Japanese) {\url http://www.c-technol.co.jp/cms/wp-content/uploads/2014/04/447fbn.pdf}
 
\bibitem{naitoevaluation2014} W. Naito, M. Uesaka, C. Yamada, and H. Ishii, ``Evaluation of dose from external irradiation for individuals living in areas affected by the Fukushima daiichi nuclear plant accident'', Radiation Protection Dosimetry 163 (2015), 10.1093/rpd/ncu201. 

\bibitem{airborne} Airborne Monitoring Survey Results are published on the Nuclear Regulation Authority web site \url{http://radioactivity.nsr.go.jp/en/list/307/list-1.html}, last accessed May 10, 2015.

\bibitem{mapurl} The geological society of Japan web site (in Japanese) \url{http://www.geosociety.jp/hazard/content0058.html}, last accessed May 10, 2015.

\bibitem{minatoterrestrial2010} S. Minato,``Terrestrial gamma ray dose rate map of the Japanese islands in relation to geological systems'', RADIOISOTOPES 55, 543-548 (2010).

\bibitem{dshuttle} The technical specifications for D-shuttle are available at \url{http://www.c-technol.co.jp/eng/e-dshuttle}, last accessed May 10, 2015.

\bibitem{irsnmonitor} IRSN ambient dose monitor, available at \url{http://sws.irsn.fr/sws/mesure/index}, last accessed May 10, 2015. (Bastia = 102, Poitiers 84, Boulogne 64)

\bibitem{polandatlas} Isajenko K,Piotrowska B, Fujak M and Kardaś M 2011 Radiation atlas of Poland, central laboratory for radiological protection {\url http://www.gios.gov.pl/images/dokumenty/pms/monitoring\_promieniowania\_jonizujscego/Atlas\_Radiologiczny\_Polski\_2011.pdf}.

\bibitem{belarusdose} Radiation situation in Belarus, online data available at \url{http://rad.org.by/radiation-belarus.html}, last accessed May 10, 2015.

\bibitem{iso4037} ISO 4037-3: ``X and gamma reference radiation for calibrating dosemeters and doserate meters and for determining their response as a function of photon energy - Part3: Calibration of area and personal dosemeters and the measurement of their response as a function of energy and angle of incidence''.

\bibitem{icrp103} ICRP, 2007. The 2007 Recommendations of the International Commission on Radiological Protection. ICRP Publication 103. Ann. ICRP 37 (2-4).

\bibitem{soildatabase} Data can be downloaded from the web site: Geochemical map of sea and land of Japan, \url{https://gbank.gsj.jp/geochemmap/data/data.htm}, last accessed May 10, 2015.

\bibitem{beck} H.L. Beck,  J. DeCampo,  and C. Gogolak, ``In 
 situ  Ge(Li)  and  NaI(Tl)  gamma-ray  spectrometry''.  USAEC  Report  HASL-258,  New  York,  N.Y. 
 10014  (1972). 

\bibitem{icrp74} ICRP, 1996. Conversion Coefficients for use in Radiological Protection against External Radiation. ICRP Publication 74. Ann. ICRP 26 (3-4).

\bibitem{why033} R. Ichikawa, ``Estimate of National Dose of Radiation due to the Living Environment in Japan'' (in Japanese), RADIOISOTOPES  62 (2013) 927-938, and references therein.

\bibitem{unscear2000} UNSCEAR 2000 REPORT: Sources and Effects of Ionizing Radiation,  United Nations Publication, ISBN 92-1-142238-8.

\bibitem{frenchdose} Rannou A, Aubert B and Scanff P 2006 Exposition de la population française aux rayonnements
ionisants Rapport IRSN DRPH/SER 2006–02.

\bibitem{ICRP111} ICRP, 2009. Application of the Commission's Recommendations to the Protection of People Living in Long-term Contaminated Areas After a Nuclear Accident or a Radiation Emergency. ICRP Publication 111. Ann. ICRP 39 (3).



\end{thebibliography}
\end{document}